\documentstyle[preprint,prb,aps,psfig]{revtex}

\begin{document}

\title{Reevaluating electron--phonon coupling strengths:
Indium as a test case for {\it ab initio} and many-body--theory methods}

\author{Sven P. Rudin$^\dag$, R. Bauer,$^\ddag$ Amy Y. Liu,$^\dag$ and
J. K. Freericks$^\dag$}

\address{
$^{\dag}$ Department of Physics, Georgetown University,
Washington, D.C. 20057-0995, U.S.A.\\
$^{\ddag}$ Theoretische Physik, Universit\"at Regensburg,
D-93040 Regensburg, Germany}

\maketitle

\baselineskip22pt

\begin{abstract}
\baselineskip22pt
Using indium as a test case, we investigate the accuracy of the
electron--phonon coupling calculated with state-of-the-art {\it ab initio}
and many-body theory methods.  The {\it ab initio} calculations ---
where electrons are treated in the local-density approximation,
and phonons and the electron--phonon interaction are treated within
linear response --- predict an electron--phonon spectral function
$\alpha^2 F(\omega)$ which translates into a relative tunneling
conductance that agrees with experiment to within one part in 10$^{3}$.
The many-body theory calculations --- where $\alpha^2 F(\omega)$ is
extracted from tunneling data by means of the McMillan-Rowell tunneling
inversion method --- provide spectral functions that depend strongly on
details of the inversion process. For the the most important moment of
$\alpha^2 F(\omega)$, the mass-renormalization parameter $\lambda$,
we report 0.9$\pm0.1$, in contrast to the value 0.805 quoted for nearly
three decades in the literature. The {\it ab initio} calculations also
provide the transport electron--phonon spectral function
$\alpha_{tr}^2 F(\omega)$, from which we calculate the resistivity
as a function of temperature in good agreement with experiment.
\end{abstract}

\vskip1.0in


\vfill

\section{Introduction}

Materials with phonon-mediated superconductivity were most intensively
studied two to three decades ago.
In the last decade, especially since the discovery of compounds with
high transition temperatures, the experimental study of such low-temperature
materials has steadily declined.
In contrast, recent years have seen a steady improvement in computational
and theoretical methods aimed at describing the electron-phonon
coupling in the ``old'' materials.
State-of-the-art {\it ab initio} methods can now be used to study details
of the interaction between electrons and phonons and to estimate 
transition temperatures.
The accuracy of these calculations, in itself worth investigating, also
raises the question of how precisely we know the parameters derived from
experiments.

The understanding of phonon-mediated superconductivity relies on a
detailed description of the coupling between phonons and electrons,
most explicit in the electron--phonon spectral function
$\alpha^2 F(\omega)$ of Eliashberg theory.\cite{eliashberg60}
The spectral function measures the strength with which phonons scatter
electrons on the Fermi surface with an imparted energy $\omega$.
With the addition of an effective Coulomb repulsion, i.e., the
Morel-Anderson pseudopotential $\mu^*$, $\alpha^2 F(\omega)$ determines
all the thermodynamic properties of a phonon-mediated superconductor,
including the transition temperature $T_C$, the critical field, and the
specific heat jump at $T_C$.\cite{carbotte90}
Rather than treating all scattering events equally, one can also
weight each scattering event according to how much the direction of
the electronic velocity changes.  
This weighting results in the {\it transport} electron--phonon spectral
function $\alpha_{tr}^2 F(\omega)$, which determines the transport
properties in the normal state.

First-principles density-functional calculations  can be used to study
the electronic structure, vibrational properties, and electron-phonon
coupling in real materials. 
To calculate quantities such as $\alpha^2 F(\omega)$ and
$\alpha_{tr}^2 F(\omega)$, which involve averages over all phonon modes,
the density-functional linear-response approach is particularly 
useful.\cite{giannozzi91,quong92,savrasov92,savrasov94,liu96}
In this approach, the electronic response to atomic displacements is
determined self-consistently, and phonon wave vectors throughout the
Brillouin zone are accessible without having to construct the large
supercells needed in finite-difference-based frozen-phonon or
generalized supercell methods.\cite{lam82,weichou,heid98}
This approach has been successfully used to study the electron-phonon
interaction and superconductivity in many simple metals that are
suitably treated by the approximations inherent in the method, i.e.,
the local density approximation (LDA) for the electrons and the harmonic
approximation for the phonons.\cite{savrasov94,liu96,bauer98}

Alternatively, $\alpha^2 F(\omega)$ and $\alpha_{tr}^2 F(\omega)$ can
be extracted from tunneling experiments and optical conductivity data,
respectively,\cite{mcmillan65,allen71} using the Migdal-Eliashberg
theory of superconductivity. 
In particular for $\alpha^2 F(\omega)$, structure in the tunneling
conductance  measured across metal-insulator-superconductor junctions
reflects structure in the superconducting gap function $\Delta(E)$
resulting from the interaction of the electrons with the
phonons.\cite{schrieffer63,wolf85}
In the McMillan-Rowell tunneling inversion method,\cite{mcmillan65}
the Eliashberg equations are solved iteratively to find an
$\alpha^2 F(\omega)$ that is consistent with the measured
tunneling spectrum. 
In recent years, improved computational strategies for solving the
Eliashberg equations have been developed, allowing for more accurate
extractions of $\alpha^2 F(\omega)$ from tunneling data.\cite{freericks97}

In this contribution, we focus on $\alpha^2 F(\omega)$ determined from
{\it ab initio} and many-body theory methods to investigate the accuracy
with which electron--phonon parameters are known.
The moments of $\alpha^2 F(\omega)$, e.g., the electron--phonon
mass-renormalization parameter $\lambda$, are quoted for many materials
with several digits.
Our results show that this is misleading, since for indium we find
that $\lambda$ can only be given to within approximately 10\%.

Indium is an ideal candidate for our discussion  because high-quality
tunneling data are available.
It is possible to fabricate clean tunnel junctions and the
electron-phonon coupling strength appears large enough to yield a good
signal-to-noise ratio.\cite{dynesxx,dynes70}  
From the theoretical point of view, indium also serves as a good test
case, since it is a relatively simple metal in which relativistic
effects are small. 
The core--valence interaction can be accurately treated with a
pseudopotential,  and since there are only $s$ and $p$ valence electrons,
the electronic wave functions can be expanded efficiently in plane waves.  
Furthermore, our results indicate that the structural and electronic 
properties of indium are well described by the LDA and that anharmonic
effects are small at temperatures near or below the Debye temperature. 

{\it Structure of this paper.}
The first-principles and many-body theory calculations are outlined 
in Section~II.  We refer readers to the references for discussions
of the many finer, technical points not included here.
We present and discuss the results of our calculations in Section~III,
and give concluding remarks in Section~IV.

\section{Outline of the Calculations}

The {\it ab initio} procedure consists of three parts: the electronic
structure, the vibrational properties, and the electron--phonon coupling.

The electronic structure is calculated in the local density approximation
(LDA) of density functional theory, by solving the Kohn-Sham equations
self-consistently using the Perdew-Zunger parameterization of the
correlation energy.\cite{perdew81}
Since it is primarily the valence electrons that determine the structure
and hence interact with the phonons, the core electrons are eliminated from
the calculation by using a pseudopotential, which is generated by
the improved Troullier and Martins scheme.\cite{troullier91}
The nonlinearity of the exchange and correlation interaction between
the core and valence charge densities is handled with partial core
corrections.\cite{louie82}
The Kohn-Sham orbitals are expanded in plane waves with a kinetic
energy cutoff of 20~Ry.

Integrations over the Brillouin zone are approximated by sums over
discrete sets of {\bf k}-points.
The Kohn-Sham orbitals are calculated for 1056 {\bf k}-points in the
irreducible Brillouin zone (IBZ).
These points, generated with the Monkhorst-Pack scheme,\cite{monkhorst76}
originate from a mesh of 24$^3$ {\bf k}-points in the full Brillouin zone.
To accelerate convergence for this metallic system, we  use first-order 
Hermite-Gaussian smearing with a width of 0.04~Ry.\cite{methfessel89} 
The electronic density of states, in particular the density of states at
the Fermi level, $N(E_F)$, is calculated more accurately using the
linear tetrahedron method.\cite{lehman72}

The vibrational properties are determined by calculating the
self-consistent first-order change in the electron density with respect
to atomic displacements.\cite{quong92}
For each phonon wave vector {\bf q}, this change is used to calculate
the dynamical matrix, which in turn is diagonalized to give the phonon
eigenvectors ${\bf e}_{{\bf q}\nu}$ and frequencies
$\omega_{{\bf q}\nu}$ ($\nu$ is a branch label). 

Because the linear-response calculation is the most time-consuming step
in the {\it ab initio} procedure, we calculate the phonons for a
relatively small set of 59 {\bf q}-points in the IBZ.
The dynamical matrices are obtained on a finer mesh of {\bf q}-points
by a Fourier deconvolution, where the calculated dynamical matrices are
Fourier-transformed to obtain the real-space force constants, which 
can then be used to form the dynamical matrix at arbitrary {\bf q}-points.
The phonon dispersion in the vicinity of {\bf q}=0 is found to be
sensitive to the number of atomic shells included in the force constant model.
To ensure the accuracy of the long-range force constants, we also do the
full linear-response calculation for several small {\bf q}-points not in
our original mesh of 59 points. 

The final {\it ab initio} step is to calculate the coupling of each phonon
to the electron states.
A phonon ${\bf q}\nu$ will scatter an electron from a state
$|n{\bf k}\rangle$ to a new state $|n'{\bf k}'\rangle$ with a strength
determined by the resulting first-order change $\delta V_{{\bf q}}^{SCF}$
in the self-consistent potential.
For atoms of mass $M$, the electron--phonon matrix elements are given by
\begin{equation}
g (n{\bf k}, n'{\bf k}', \nu{\bf q}) =
\sqrt{\frac{\hbar}{2M\omega_{{\bf q}\nu}}}
\langle n'{\bf k}'|
{\bf e}_{{\bf q}\nu}
\cdot
\delta V_{{\bf q}}^{{\bf SCF}}
| n{\bf k} \rangle
, \label{eq:elphmx}
\end{equation}
with the restriction ${\bf k}'={\bf k}+{\bf q}$.
Since only electrons near the Fermi surface can scatter via phonons, the
average coupling  of electrons to a phonon ${\bf q}\nu$ is expressed in the
doubly-constrained Fermi surface average,
$\langle\langle |g_{{\bf q}\nu}|^2 \rangle\rangle$.\cite{lam82}
As with the dynamical matrices, the electron--phonon matrices are
calculated on the coarse mesh of 59 {\bf q}-points and then interpolated
to a denser mesh by means of a Fourier deconvolution.
The electron--phonon spectral function, which involves coupling to
all phonons, is given by
\cite{allen72}
\begin{equation}
\alpha^2 F(\omega) = N(E_F) \sum_{{\bf q}\nu}
\;
\delta(\hbar\omega - \hbar\omega_{{\bf q}\nu})
\langle\langle |g_{{\bf q}\nu}|^2 \rangle\rangle
. \label{eq:a2f}
\end{equation}
In our calculations, the $\delta$ function in Eq. (\ref{eq:a2f})  is 
replaced by a Gaussian of width 0.5~meV.

With slight modifications, the above formalism can be used to compute
the transport spectral function $\alpha_{tr}^2 F(\omega)$ for the
phonon-limited electrical resistivity, $\rho(T)$, where not all
scattering events are equally important.
For example, forward scattering events do not change the direction of the
electron velocity ${\bf v}_{n{\bf k}}$ and do not contribute to
the resistivity. 
To this end an efficiency factor,\cite{allen71}
\begin{equation}
\eta_{n{\bf k}, n'{\bf k}' }
= 1 - \frac{{\bf v}_{n{\bf k}}
      \cdot {\bf v}_{n'{\bf k}'}}
           {| {\bf v}_{n{\bf k}} |^2 }
, \label{eq:effi}
\end{equation}
is used to weight the electron--phonon matrix elements in the calculations
of  $\alpha_{tr}^2 F(\omega)$.   
The electrical resistivity is then given by\cite{allen71,allen78}
\begin{equation}
\rho(T) = \frac{3\pi\Omega}{e^2 N(E_F)
\langle {\bf v}^2 \rangle}
\frac{1}{2k_B T}
\int_{0}^\infty \hbar\omega
\frac{\alpha_{tr}^2 F(\omega)}{\sinh^2 (\hbar\omega/2k_B T)}
d\omega
,
 \label{eq:resi}
\end{equation}
with $\Omega$ the cell volume and $\langle {\bf v}^2 \rangle$ the
Fermi-surface average of the electron velocity.
This is a variational solution to the semiclassical Boltzmann equation in
which the Fermi surface is assumed to undergo a uniform rigid shift in an
applied electric field. 
Here current--current vertex corrections are included via the efficiency
factor, but only to lowest order. 
As written, Eqs. (\ref{eq:effi}) and (\ref{eq:resi}) are appropriate for
isotropic and nearly isotropic materials since $\alpha_{tr}^2 F(\omega)$
is averaged over all directions  and $\langle v_x^2 \rangle$
is assumed to be equal to $\langle {\bf v}^2 \rangle/3$. 

We compare the {\it ab initio} electron--phonon coupling with experiment
both by calculating the tunneling conductance from the {\it ab initio}
$\alpha^2 F(\omega)$ and by extracting the experimental
$\alpha^2 F(\omega)$ from tunneling data.
The traditional procedure, for which the new computational strategies
have been developed, is to calculate the experimental
$\alpha^2 F(\omega)$ from tunneling data by solving the Eliashberg
gap equations.

The extraction of $\alpha^2 F(\omega)$ from tunneling data is done with the
McMillan and Rowell tunneling inversion method.\cite{mcmillan65}
We follow their original prescription:
(i) We assume an initial value for $\alpha^2 F(\omega)$ for which
(ii) we adjust $\mu^*$ to reproduce the experimental
superconducting gap at zero temperature $\Delta_0$,
which is 0.541~meV for indium;
(iii) we compute the functional derivative of the change in the tunneling
density of states with respect to a change in the assumed
$\alpha^2 F(\omega)$;
(iv) we determine the required
shift in $\alpha^2 F(\omega)$ to produce the experimental tunneling
conductance via a singular-value-decomposition, and
(v) we determine the new $\alpha^2 F(\omega)$ by adding a smoothed shift
$\delta \alpha^2 F(\omega)$ to it.
The new $\alpha^2 F(\omega)$ is then used to begin again with step (ii),
and the entire process is repeated until it converges.

While the McMillan--Rowell tunneling inversion procedure is well defined,
different computational strategies can be used to solve the Eliashberg
equations. 
We perform the perturbation theory directly on the imaginary-frequency
axis with an energy cutoff of six times the maximum phonon frequency,
$\omega_{max}$ (beyond which $\alpha^2 F(\omega)=0$), and then perform an
exact analytic continuation to the real axis.\cite{freericks97,marsiglio88}
This method treats the Morel-Anderson pseudopotential properly because
the sharp cutoff on the imaginary-frequency axis translates into a smooth
cutoff when analytically continued to the real axis.\cite{leavens80}
In addition, the perturbation theory is performed relative to the exact
result in the normal state.
These details are necessary to accurately predict a superconducting $T_C$
from the experimental data.
The inputs are the experimental tunneling conductance and the superconducting
gap at zero temperature.
The outputs are the electron--phonon spectral function
$\alpha^2 F(\omega)$ and the Morel-Anderson pseudopotential $\mu^*$.
The transition temperature $T_C$ is then calculated with the T-matrix
method of Owen and Scalapino,\cite{owen71} rather than using
an approximate equation such as the McMillan formula.

The accuracy of the tunneling experiments suffers at low energy,
where the signal near the gap edge shows a large slope, and at
high energy, where the detailed structure in $\Delta(E)$ is washed out
because it enters the measurement in the form $E^2-\Delta^2(E)$.
The experimental data also depends critically on the precise value of
the superconducting gap at zero temperature, $\Delta_0$, because this
produces the BCS form for the tunneling conductance;
the strong-coupling corrections,
which are employed to extract $\alpha^2 F(\omega)$,
are the deviations from the BCS form.
Hence an $\alpha^2 F(\omega)$ extracted from tunneling data is usually
constrained by assuming that it has a quadratic dependence at low energy
and that it vanishes beyond a maximal phonon frequency
(some researchers include a quadratic dependence at high energies too).
Unfortunately, neither the upper limit of the low-energy quadratic behavior,
nor the exact value of $\omega_{max}$ or the frequency dependence near
the $\omega_{max}$ is known.
In principle, $\omega_{max}$ should be chosen to be equal to the maximum
bulk phonon frequency, but it frequently is allowed to be somewhat larger
to allow for the effects of interface phonons.
We adjust the region of quadratic behavior and the maximal phonon frequency
in different fitting procedures, and impose a linear form on
$\alpha^2 F(\omega)$ to bring it to zero at $\omega_{max}$.

\section{Results and Discussion}

The ground-state crystal structure of indium is face-centered tetragonal
(fct).
As is typical with LDA calculations,  we find that compared to experiment
the equilibrium volume is approximately 5\% too small.
The tetragonal lattice parameters are calculated to be $a$=4.51~\AA~ and
$c$=4.84~\AA; the experimental values are $a$=4.58~\AA~ and
$c$=4.94~\AA.\cite{smith69}
The calculated and measured $c/a$ ratios agree to better than 1\%. 
Our linear-response calculations are all performed using the theoretical
lattice parameters.

The calculated electronic density of states (DOS) is plotted in Figure 1.
The DOS has a free-electron-like behavior at low energies, but develops
more structure at higher energies where bands cross the Bragg planes.
The dashed curve shows the DOS for a free electron gas with the same 
average valence-electron density as indium.
The two curves differ significantly, indicating that band-structure
and correlation effects within LDA
strongly renormalize the electron mass in indium.
The DOS at $E_F$ is reduced by about 26\% compared to the free
electron value. 
Unfortunately, it is difficult to compare the calculated DOS directly with
magnetic susceptibility measurements for indium since these experiments
find diamagnetic rather than Pauli paramagnetic behavior at low
temperatures.

The inset to Figure 1 shows the DOS near $E_F$, where the DOS varies
only by a few percent on the scale of phonon energies.
This is in particular true over the range $E_F \pm 6 \omega_{max}$,
where we assume a constant electronic DOS for the tunneling inversion.

Figure 2 shows the excellent agreement between the measured and calculated
phonon dispersion curves.
The experimental data is taken from neutron diffraction, and is reported
with an 11th-neighbor, 19-parameter Born--von K\'arm\'an
fit.\cite{smith69}
Along the direction from Z to X for which no direct experimental data
are available, we find good agreement between the fit to the experiment
and our calculated phonon dispersion, though the latter shows more
structure than the fit.
This structure, if real, may be more detailed than can be extracted from
the available experimental data.

The electrical resistivity, calculated with Eq. (\ref{eq:resi}),
is plotted along with experimental data from polycrystalline
samples\cite{resist} in Figure 3.
Eq. (\ref{eq:resi}) is expected to be most accurate in the temperature
range  of about
$\Theta_D/5 \mathrel{\raise.3ex\hbox{$<$\kern-.75em\lower1ex\hbox{$\sim$}}}
T \mathrel{\raise.3ex\hbox{$<$\kern-.75em\lower1ex\hbox{$\sim$}}}
2\Theta_D$,
with the Debye temperature $\Theta_D =$129~K.\cite{ashcroft76}
At very low temperatures anisotropy effects become important, while
at high temperatures anharmonic effects must be included.\cite{pinski81}
Although the crystal structure of indium is tetragonal, electrical
resistivity measurements on single crystals find nearly the same results
along the $a$ and $c$ directions. 
This isotropy also appears in our calculations, where
$\langle v_x^2 \rangle$ and $\langle v_z^2 \rangle$ differ by less than 5\%. 

The dashed curve in Figure 3 is calculated using $\alpha^2 F(\omega)$,
which is often used as an approximation to $\alpha_{tr}^2 F(\omega)$.
The rough agreement between the dashed curve and the measured resistivity
justifies the approximation in cases when $\alpha_{tr}^2 F(\omega)$ is
not known. 
However, including the correct weighting with the efficiency factor 
$ \eta_{n{\bf k}, n'{\bf k}' } $
brings the calculated resistivity into much better agreement with experiment
for temperatures up to well above the Debye temperature.
The transport electron-phonon coupling parameter $\lambda_{tr}$ is found
to be 0.74, as compared to $\lambda=0.88$.
The effect of the efficiency factor on the spectral function is shown in
the inset of Figure 3, where we also display $F(\omega)$ scaled to
emphasize the strikingly similar shape of the phonon density of states
and the electron-phonon spectral functions.
The ratio of $\alpha^2 F(\omega)$ to $F(\omega)$ gradually increases
with frequency at about one third of the rate seen in lead.\cite{carbotte90}

Figure 4 shows four electron--phonon coupling functions $\alpha^2 F(\omega)$:
The calculated {\it ab initio} result and three curves extracted from
experimental tunneling data.
All three extracted curves are based on the same tunneling data taken
at $T=0.35$~K,\cite{dynesxx}
which should be more accurate than the data taken at higher
temperatures.\cite{dynes70}
The extracted curves differ in the constraints imposed on their low- and
high-frequency behavior.
The unconstrained curve is quadratic for $\omega < 0.5$~meV and uses
$\omega_{max}=21$~meV.
While this curve yields the best fit to the relative tunneling conductance,
it shows what is believed to be unphysical behavior at low and high energies.
There is no reason to believe that there is a phonon feature at
1.5~meV as shown in the unconstrained curve.
Rather, that shoulder is most likely an artifact related to the accuracy
of the voltage (and of $\Delta_0$) for the experimental data collected
at low energies.
The features at high energy may be real, i.e., structure from either
vertex corrections or from interface phonons, but most likely they arise
from forcing an accurate fit to the experimental data at approximately
13~meV above the superconducting gap. 
We estimate that vertex corrections lead to a small reduction of $T_C$
of approximately 0.3\%, based on a simple integral of
$\alpha^2 F(\omega)$
(using the Fermi-surface average $C=0.18$).\cite{freericks97}
This result is the same size of effect as seen in lead, so vertex
corrections can be safely neglected for indium.

The constrained curve is more strongly restricted in its shape at both
low- and high-energies:
$\alpha^2 F(\omega)$ is forced to increase quadratically in $\omega$
for $\omega < 2$~meV and decay linearly to zero at the maximum bulk
phonon frequency of 16~meV.
These constraints eliminate what appear to be unphysical features in the
unconstrained $\alpha^2 F(\omega)$, while still fitting the tunneling data
extremely well. 
The fourth curve shown in Figure 4 is that of Dynes,\cite{dynesxx}
on which the three-decade-old value for $\lambda$ is based.
His calculation differs from ours in that it was performed directly on
the real axis, which does not handle $\mu^*$ properly,\cite{leavens80}
and it was not performed relative to the normal state, which would enforce
the correct energy dependence at high energies.

For each of the $\alpha^2 F(\omega)$ shown in Figure 4,
the differences between the measured and calculated tunneling
conductance are
plotted in Figure 5.
It is remarkable that an {\it ab initio} calculation with only
one adjustable parameter
(the Morel-Anderson pseudopotential $\mu^*$, adjusted to give the
superconducting gap at zero temperature)
can fit the experimental tunneling conductance
to better than one part in 10$^3$
(the tunneling conductance is on the order of 1).
Furthermore, we see that the low- and high-frequency features 
unique to the
unconstrained curve greatly improve the fit only in the low and high
energy ranges.
Since the experimental data is least accurate in these ranges,
it makes sense to
constrain the fitting procedure to suppress the unphysical features
that stem from these ranges.
Ideally, the best way to proceed would be 
to use experimental data that has error bars reported with it.
Such data
would allow a maximum entropy technique to be employed to produce the 
best fit $\alpha^2 F(\omega)$.\cite{jarrell96}

Table~I describes the curves extracted from tunneling data and the
{\it ab initio} calculation.
The electron-phonon spectral functions are characterized by
several moments:\cite{carbotte90}
(i)
the electron--phonon mass-renormalization parameter $\lambda$
(twice the first inverse moment),
(ii)
the strength A
(the area under the curve), and
(iii) the characteristic phonon energy $\omega_{ln}$
(a logarithmic moment).
The extracted $\alpha^2 F(\omega)$ and Morel-Anderson pseudopotential
$\mu^*$ (adjusted to
reproduce the experimental $\Delta_0$) are employed to calculate
the critical temperature $T_C$ with no further adjusting of parameters.
All results lie within 5\% of the experimental $T_C$.
The errors
in the tunneling conductance are all rather close
(with exception of the unconstrained curve),
whereas the curves differ significantly in their moments.

One possible explanation for the wide variation in the moments is
the smallness of $\lambda$ for indium combined with
the experimental uncertainty in the superconducting gap at zero
temperature, $\Delta_0$.
Since $\mu^*$ is adjusted to give $\Delta_0$ and materials with small
$\lambda$ do not display strong features in the tunneling DOS,
it is difficult to extract $\alpha^2 F(\omega)$ to high accuracy.
The value of $\alpha^2 F(\omega)$ in the region between 0 and 3~meV has
a large effect on the size of the extracted $\lambda$, but this is the
region where the experimental data depends most on the precise knowledge
of $\Delta_0$ and the experimental voltage.

Given the range of $\lambda$ from
0.8 to 1.1 for the different $\alpha^2 F(\omega)$, it is not surprising that
$\mu^*$ also spans a wide range.   Conventional wisdom
limits $\mu^*$ to the range of 0.1 to 0.14 for
most materials.  In fact, this was a criterion used for
choosing junctions in the tunneling experiments.\cite{dynes70}
However, our {\it ab initio} $\alpha^2 F(\omega)$ as well as our constrained
and unconstrained
$\alpha^2 F(\omega)$ extracted from tunneling data all give $\mu^*$
larger than the conventional values.  
Recent first-principles calculations of  $\mu^*$
suggest that $\mu^* < 0.14$ is an artificial limit for
some simple metals.\cite{Jin}
Within the standard Eliashberg theory, where a
constant electronic DOS is assumed, 
$\mu^*=\mu / [1-\mu\ln ( N( E_F )
	\, 6 \omega_{max} )]$,
and the maximum value is found by letting
$\mu$ become infinite.   For indium, this gives
a maximum $\mu^*$ of about 0.25.
We get the same estimate for a maximal $\mu^*$ by including
the energy dependence of the electronic DOS,
i.e., the $ N(E) $ of Fig.~1, in\cite{marsiglio92,freericks95}
\begin{equation}
\mu_{max}^* =
\frac{\pi N(E_F)}
{\int_{-\infty}^\infty \frac{dy}{y}
N(E_F + y)
\left(
\tan^{-1} \left(
\frac{y}{6 \omega_{max}}
\right)
-
\tan^{-1} \left(
\frac{y}{\omega_p}
\right)\right)
}
, \label{eq:muest}
\end{equation}
with a plasma frequency $\omega_p \approx 12$~eV.
We expect $\mu < 1$ in indium because it is an $s$-$p$ metal,
well described by the free-electron model,
so expected values of $\mu^*$ should be less than 0.2.
All these estimates indicate that the unconstrained curve is unphysical.

Even if we discount the unconstrained curve, the remaining
values for $\lambda$ differ by up to 20\%. 
Low-temperature specific heat data
can be used to provide an additional estimate for $\lambda$.
Using the linear coefficient $\gamma$ from experiment\cite{oneal65}
and our calculated electronic DOS at the Fermi level,
we estimate $\lambda = 0.86$, which is close to the {\it ab initio}
value.
This estimate is uncertain because the experimental $\gamma$ itself
is known only to a few percent,\cite{oneal65} and furthermore the estimate
relies on a precise knowledge of
the electronic DOS at the Fermi level and on the assumption
that electron--electron effects do not contribute significantly
to the mass renormalization.
Taken together, these results lead us to conclude that
for indium $\lambda = 0.9 \pm 0.1$.

\section{Conclusions}

State-of-the-art {\it ab initio} methods deliver
a very accurate description of the electron--phonon coupling in indium:
The calculated relative tunneling conductance
agrees with experiment to better than one part in 10$^3$;
the calculated intrinsic resistivity as a function of
temperature is also in good agreement with experiment.
The achieved accuracy justifies the approximations invoked:
The local density approximation used to calculate the electronic structure
and the harmonic approximation for the phonons. 

Indium is a good choice for the comparison also because of the high
quality experimental tunneling data.
Still, we do not know the strength of the electron--phonon
mass-renormalization parameter $\lambda$ as well as it would
seem from the literature.
Based on our calculations with state-of-the-art many-body theory methods,
we estimate that $\lambda$ can only be determined to within 10\%,
because of uncertainties in the data at low and high energies.
The uncertainties lead to
the question of how to best extract the electron--phonon
spectral function $\alpha^2 F(\omega)$ from experimental data:
Is it better to fit
the data as precisely as possible or to allow for experimental
errors at low and high energies by constraining the curve to be
physically reasonable?
All the $\alpha^2 F(\omega)$ --- {\it ab initio}, many-body with and
without constraints --- show the same structure with roughly the same
magnitude.
The tunneling conductance obtained from our {\it ab initio}
$\alpha^2 F(\omega)$ is as accurate as the tunneling conductance obtained
from the most likely spectral function extracted from the experimental data.

From our study we conclude that the accuracy with which the electron--phonon
coupling strength is extracted from experiment could be improved.
In particular, we hope to motivate further experimental
work that reports error bars for the tunneling conductance and
the superconducting gap
so that a maximum-entropy technique can be employed
to determine the best fit $\alpha^2 F(\omega)$.
In materials where vertex corrections are more important,
the improved accuracy of $\alpha^2 F(\omega)$ would also allow the effects of
vertex corrections to be observed in the multiphonon region.

\bigskip

\centerline{ACKNOWLEDGMENTS}

We acknowledge useful discussions with
P.~B.~Allen,
R.~Dynes,
M.~Mihjak,
P.~Miller,
E.~Nicol,
A.~Quong,
J.~Rowell,
J.~W.~Wilkins,
and
V.~Zlati\'c.
R.~Bauer acknowledges support from the DAAD.
This work was supported by the National Science Foundation under Grant
DMR-9627778 and
by the Pittsburgh Supercomputing Center under Grant DMR970008.

\begin{figure}
  \caption{
Electronic density of states from the {\it ab initio} calculation; inset:
density of states near the Fermi level.
The {\it ab initio} results (solid curve)
are compared to the free-electron density of states
(dashed curve).
The inset shows that the calculated density of states changes by only a
few percent within the range of
$E_F \pm 6 \omega_{max}$,
where $\omega_{max}$ is the maximum
phonon frequency;
this justifies the assumption of a constant DOS in this region
for the many-body-theory calculations.
   }
\label{fig:1}
\end{figure}

\begin{figure}
  \caption{
Phonon dispersion along the high-symmetry directions
shown in the inset.
The experimental data and corresponding fit are from Ref.~30.
The {\it ab initio} results are calculated on a uniform grid of 59
{\bf q}-points in the irreducible wedge of the BZ and then interpolated
to a denser mesh by means of a Fourier deconvolution.
   }
\label{fig:2}
\end{figure}

\begin{figure}
  \caption{
Electrical resistivity as a function of temperature.
The dashed and solid curves are calculated using the {\it ab initio}
results for the electron-phonon spectral function $\alpha^2 F(\omega)$ and
its transport analog $\alpha_{tr}^2 F(\omega)$, respectively.   The latter
agrees significantly better with
the experimental data.\cite{resist}
The inset shows the strikingly similar shape
of the phonon density of states $F(\omega)$,
$\alpha^2 F(\omega)$, and $\alpha_{tr}^2 F(\omega)$,
with the units scaled to display $F(\omega)$ with a magnitude similar
to that of the spectral functions.
   }
\label{fig:3}
\end{figure}

\begin{figure}
  \caption{
Electron--phonon spectral functions $\alpha^2 F(\omega)$
from {\it ab initio} calculations and
extracted from experimental tunneling results.
The three extracted curves are all based on the
same tunneling data, but differ in
the constraints used in the tunneling inversion procedure.
The unconstrained curve shows spurious behavior at low
frequencies and above the maximum phonon frequency.
   }
\label{fig:4}
\end{figure}

\begin{figure}
  \caption{
Differences between the measured and calculated tunneling conductances
based on the spectral functions $\alpha^2 F(\omega)$
in Fig. 4. The unconstrained $\alpha^2 F(\omega)$ yields
errors in the conductance that are two orders of magnitude smaller than
any of the other $\alpha^2 F(\omega)$ curves.
   }
\label{fig:5}
\end{figure}

\begin{table}
\begin{tabular} {rcccccccc}
&\multispan3{\hfil}&
\multispan2{\hfil}&
\multispan2{error in tunneling conductance}\\
&$\lambda$&A (meV)&$\omega_{ln}$ (meV)&$\mu^*$&$T_C$ (K)
&max ($10^{-3}$)&rms ($10^{-4}$)\\
\tableline
experiment\cite{expIn}&&&&&3.40\\
\tableline
{\it ab initio}&0.882&3.00&5.61&0.161&3.31&1.1&6\\
unconstrained&1.108&3.66&5.20&0.326&3.24&0.04&0.08\\
constrained&0.984&3.24&5.51&0.224&3.28&2.3&5\\
Ref. 17&0.805&2.74&5.84&0.119&3.32&0.9&7\\
\end{tabular}
  \vskip0.6in
\caption{Parametric description of the $\alpha^2 F(\omega)$ functions
from the {\it ab initio} calculation and extracted from tunneling data.
The electron-phonon spectral functions are characterized by
several moments:
(i)
the electron--phonon mass-renormalization parameter $\lambda$
(twice the first inverse frequency moment),
(ii)
the strength A (the area under the curve), and
(iii) the characteristic phonon energy $\omega_{ln}$
(a logarithmic moment).
The Morel-Anderson pseudopotential $\mu^*$ is adjusted to
reproduce the experimental gap at $T$=0.  Results
for the critical temperature $T_C$,
calculated from $\alpha^2 F(\omega)$ and $\mu^*$,
all lie within 5\% of the experimental $T_C$.
Also listed are the maximum and the RMS errors
in the tunneling conductance.
}
\end{table}

\end{document}